\documentclass[12pt,a4paper]{article}

\usepackage{amssymb,amsfonts,amsmath,amsthm,pstricks,pst-all}

\newcommand{\mh}{{\mathcal H}}
\newcommand{\znu}{{\mathbb Z^{\stackrel{\nu}{}}}}

\newcommand{\dom}{\,\mathrm{Dom}\,}
\newcommand{\kerr}{\,\mathrm{Ker}\,}
\newcommand{\supp}{\,\mathrm{supp}\,}
\newcommand{\diam}{\,\mathrm{diam}\,}
\newcommand{\bhl}{\mathcal B(\mathcal H_\Lambda)}
\newcommand{\dist}{\,\mathrm{dist}\,}
\newcommand{\spec}{\,\mathrm{Spec}\,}
\newcommand{\rr}{\mathbb R}

\newcommand{\cc}{\mathbb C}
\newcommand{\zz}{\mathbb Z}
\newcommand{\ol}{\Omega_\Lambda}
\newcommand{\olo}{\Omega_{\Lambda,0}}
\newcommand{\hlo}{H_{\Lambda,0}}
\newcommand{\hl}{H_{\Lambda}}
\newcommand{\ovl}{\overline{\Lambda}}
\newcommand{\ovh}{\overline{\mh}}
\newcommand{\ovo}{\overline{\Omega}}
\newcommand{\ov}[1]{\overline{#1}}
\newcommand{\opi}{\ov{\Phi}_{I_1}^{(r)}}
\newcommand{\opj}{\ov{\Phi}_{J_1}^{(b)}}

\newtheorem{theor}{Theorem}
\newtheorem{lemma}{Lemma}

\begin{document}

\title{Ground states in relatively bounded quantum perturbations of classical lattice systems}

\author{D.A.Yarotsky \footnote{Department of Mathematical Physics,
University College Dublin, Ireland; e-mail: yarotsky@mail.ru}
\footnote{on leave from Institute for Information Transmission
Problems, Moscow, Russia }}

\date{}
\maketitle

{\bf Abstract.} We consider ground states in relatively bounded
quantum perturbations of classical lattice models. We prove
general results about such perturbations (existence of the
spectral gap, exponential decay of truncated correlations,
analyticity of the ground state), and also prove that in
particular the AKLT model belongs to this class if viewed at large
enough scale. This immediately implies a general perturbation
theory about this model.


\section{Introduction and results}

It is generally expected that if a ground state of a quantum
lattice system is in a non-critical regime characterized by the
presence of a spectral gap and exponential decay of truncated
correlations, then the system remains in this phase under
sufficiently weak perturbations of a general form. Relevant
rigorous results are now available in the case of weak quantum
perturbations of some classical models
\cite{A,DK,KT1,KT2,KT,M1,M2,Y1,Y2}. Most of these results concern
perturbations which are bounded and small in the norm sense.
However, Kennedy and Tasaki obtained in \cite{KT2} general results
for perturbations, which are only relatively bounded, in some
special sense, w.r.t. the classical Hamiltonian. Moreover, using a
special transformation of the Hamiltonian, they applied this
perturbation theory to the dimerized AKLT model, which is a
genuinely quantum $SU(2)$-invariant model. The type of relative
boundedness they used does not, however, seem to allow an
extension of their result to the non-dimerized, fully translation
invariant case. In this paper we consider perturbations relatively
bounded in the quadratic form sense, which appears to fit
naturally in this and some other contexts. We prove general
results for gapped classical models with a simple ground state and
then apply them to the non-dimerized AKLT model.

We consider a  quantum ``spin'' system on the lattice $\znu$.
Throughout the paper we consider only translation invariant
interactions. Each site $x\in\znu$ is equipped with a Hilbert
space $\mh_x$, possibly infinite-dimensional. In the sequel we use
the notation $$\mh_\Lambda\equiv\otimes_{x\in\Lambda}\mh_x$$ for
Hilbert spaces, corresponding to finite subsets of the lattice. We
assume that $\mh_x$ has a preferred vector denoted $\Omega_x$. The
corresponding product state will be denoted by
$\Omega_{\Lambda,0}$: $$\Omega_{\Lambda,0}\equiv \otimes_{x\in
\Lambda}\Omega_x.$$ Also, we fix some finite set
$\Lambda_0\subset\znu$, which will be the interaction range. The
(formal) Hamiltonian has the form $$H=H_0+\Phi,$$ where $H_0$ is
the classical part and $\Phi$ the perturbation. The classical
Hamiltonian $H_0$ is given as $$H_0=\sum_{x\in\znu}h_x.$$ Here
$h_x$ is a self-adjoint, possibly unbounded operator acting on
$\mh_{\Lambda_0+x}$, where $\Lambda_0+x$ is a shift of
$\Lambda_0$. The Hamiltonian $H_0$ is classical in the following
sense. If $\mh_x$ is finite dimensional, then we assume that in
each $\mh_x$ there is an orthogonal basis containing $\Omega_x$
and such that the product basis in $\mh_{\Lambda_0+x}$
diagonalizes $h_x$. We extend in a natural way this assumption to
the case of infinite dimensional $\mh_x$ by assuming that for each
$\mh_x$ an orthogonal partition of unity, containing the
projection onto $\Omega_x$, is given, and $h_x$ is a function of
the product partition in $\mh_{\Lambda_0+x}$. Furthermore, we
assume that $\Omega_{\Lambda_0+x}$ is a non-degenerate gapped
ground state of $h_x$:
\begin{equation}\label{hx}h_x\Omega_{\Lambda_0+x,0}=0,\quad
h_x|_{\mh_{\Lambda_0+x}\ominus\Omega_{\Lambda_0+x,0}}\ge {\bf
1}.\end{equation} Now we describe the perturbation. It is given by
$$\Phi=\sum_{x\in\znu}\phi_x,$$ where $\phi_x$ is a (possibly
unbounded) symmetric quadratic form on $\mh_{\Lambda_0+x}$,
bounded relative to the quadratic form corresponding to $h_x$,
i.e. the domain of $\phi_x$ contains $\dom(h^{1/2}_x)$ and
\begin{equation}\label{phix}
|\phi_x(v,v)|\le\alpha\|h^{1/2}_xv\|^2+\beta\|v\|^2,\quad
v\in\dom(h^{1/2}_x)\end{equation} with some $\alpha,\beta$. We
assume that $\alpha<1$. The form $\phi_x$ actually need not be
closed and generated by an operator, though in all examples we
consider it is.

If $\Lambda$ is a finite volume and
$$H_{\Lambda,0}=\sum_{x:\Lambda_0+x\subset\Lambda}h_x,\quad
\Phi_\Lambda=\sum_{x:\Lambda_0+x\subset\Lambda}\phi_x,$$ then
$\Phi_\Lambda$ is again bounded relative to $H_{\Lambda,0}$ with
the same $\alpha$, because, clearly,
$\dom(H_{\Lambda,0}^{1/2})\subset\dom(\Phi_\Lambda)$ and, by
adding up (\ref{phix}),
$$|\Phi_\Lambda(v,v)|\le\alpha\|\hlo^{1/2}v\|^2+|\Lambda|\beta\|v\|^2,\quad
v\in\dom(\hlo^{1/2}).$$ It follows from the KLMN theorem that
$H_\Lambda=\hlo+\Phi_\Lambda$ is a well-defined self-adjoint
operator, defined by its quadratic form \cite{K,RS}. Throughout
the paper unbounded operators will appear only as relatively
bounded perturbations of positive operators in the quadratic form
sense, so, in order not to complicate arguments and keep the
notation simple, we will typically not distinguish between
operators and corresponding quadratic forms.

We will assume now for simplicity that $\Lambda$ is a cubic volume
with periodic boundary conditions. Clearly, in this case
$\Omega_{\Lambda,0}$ is a non-degenerate ground state of $\hlo$
with a spectral gap: $$\hlo\Omega_{\Lambda,0}=0,\quad
\hlo|_{\mh_\Lambda\ominus\Omega_{\Lambda,0}}\ge |\Lambda_0|{\bf
1}.$$ The following result is a perturbation theory for the ground
state in the case of small $\alpha$ and $\beta$.

\begin{theor}
There exist positive $\alpha$ and $\beta$, depending only on the
dimension $\nu$ and the interaction range $\Lambda_0$, such that
if condition (\ref{phix}) holds with these $\alpha,\beta$, then:

1) $H_\Lambda$ has a non-degenerate gapped ground state
$\Omega_\Lambda:$
$$H_\Lambda\Omega_\Lambda=E_\Lambda\Omega_\Lambda,$$ and for some
independent of $\Lambda$ positive $\gamma$
$$H_\Lambda|_{\mh_\Lambda\ominus\Omega_\Lambda}\ge
(E_\Lambda+\gamma){\bf 1}.$$

2) There exists a thermodynamic weak$^*$-limit of the ground
states $\Omega_\Lambda:$ $$\langle
A\Omega_\Lambda,\Omega_\Lambda\rangle
\xrightarrow{\Lambda\nearrow\znu} \omega(A),\quad
A\in\cup_{|\Lambda|<\infty}\bhl,$$ where $\mathcal B(\mh_\Lambda)$
is the algebra of bounded operators in $\mh_\Lambda$.

3) There is an exponential decay of correlations in the infinite
volume ground state $\omega:$ for some positive $c$ and
$\epsilon<1$ $$|\omega(A_1A_2)-\omega(A_1)\omega(A_2)|\le
c^{|\Lambda_1|+|\Lambda_2|}\epsilon^{\dist(\Lambda_1,\Lambda_2)}\|A_1\|\|A_2\|,\;\;
A_i\in\mathcal B(\mh_{\Lambda_i}).$$

4) If within the allowed range of perturbations the terms $\phi_x$
(or the resolvents $(h_x+\phi_x-z)^{-1}$ in the case of unbounded
perturbations) depend analytically on some parameters, then the
ground state $\omega$ is also weakly$^*$ analytic in these
parameters (i.e. for any local observable $A$ its expectation
$\omega(A)$ is analytic).

\end{theor}

{\bf Example 1} (anharmonic quantum crystal model). Let
$\mh_x=L_2(\rr^d,dq)$ and
$$H=\sum_x(-\Delta_x+V_1(q_x))+\lambda\sum_{|x-y|=1}V_2(q_x,q_y).$$
Suppose that $V_1(q)\to +\infty$ as $q\to\infty$. In this case
$-\Delta+V_1$ has a discrete spectrum with a non-degenerate ground
state. Since $-\Delta\ge 0$, we see that if for some $c_1,c_2$
$$|V_2(q_x,q_y)|\le c_1(V_1(q_x)+V_1(q_y))+c_2,\quad \forall
q_x,q_y,$$ then for sufficiently small coupling constant $\lambda$
the operator $H_\Lambda$ is well-defined by the KLMN theorem, and
Theorem 1 applies.

The next theorem extends the perturbation theory to all
$\alpha\in(0,1)$ at the cost of a slightly more stringent
assumption about the perturbation. We replace (\ref{phix}) with
the following stronger assumption:
\begin{equation}\label{p1}\phi_x=\phi^{(r)}_x+\phi^{(b)}_x,\end{equation}
where $\phi^{(r)}$ is the ``purely relatively bounded'' part of
the perturbation:
\begin{equation}\label{p2}|\phi_x^{(r)}(v,v)|\le\alpha\|h_x^{1/2}v\|^2,
\end{equation} and $\phi^{(b)}$ is the bounded part:
\begin{equation}\label{p3}\|\phi^{(b)}_x\|\le \beta.\end{equation}
In particular, (\ref{p2}) and (\ref{hx}) imply that
$\phi_x^{(r)}\Omega_{\Lambda_0+x,0}=0$ if $\phi_x^{(r)}$ is viewed
as an operator (more precisely,
$\phi_x^{(r)}(v,\Omega_{\Lambda_0+x,0})=0$ for all $
v\in\dom(h_x^{1/2})$).

\begin{theor}
For any $\varkappa>1$ there exists
$\delta=\delta(\varkappa,\nu,\Lambda_0)>0$ such that: for any
$\alpha\in (0,1)$, if conditions (\ref{p1})-(\ref{p3}) are
satisfied with this $\alpha$ and
$\beta=\delta(1-\alpha)^{\varkappa(\nu+1)}$, then all conclusions
of Theorem 1 hold.
\end{theor}

{\bf Remark.} The assumption (\ref{p2}) can be somewhat relaxed.
In fact, what is actually used in the proof of Theorem 2 is not
(\ref{p2}) but the weaker condition: for any $I\subset\Lambda$
\begin{equation}\label{wcond}\bigl|\sum_{x\in
I}\phi_x^{(r)}(v,v)\bigr|\le\alpha\|H_{\Lambda,0}^{1/2}v\|^2.\end{equation}
This is the condition which we will use when we consider the AKLT
model.

{\bf Example 2.} Consider a Hamiltonian $$H=\sum_xA_x,$$ where
$A_x$ is a self-adjoint operator on $\mh_{\Lambda_0+x}$ such that
$$A_x\Omega_{\Lambda_0+x,0}=0,\quad
A_x|_{\mh_{\Lambda_0+x}\ominus\Omega_{\Lambda_0+x,0}}\ge {\bf
1}.$$ Clearly, $\otimes_x\Omega_x$ is a ground state of $H$ with a
gap $\ge |\Lambda_0|$. We expect that a perturbation theory in the
sense of Theorems 1,2 holds at least for general weak bounded
perturbations of $H$. Theorem 2 shows that this is indeed so at
least if $\|A\|<\infty$ ($A$ is the operator whose translates
$A_x$'s are). Indeed, consider a finite range perturbation
$\sum_x\psi_x$ with small $\|\psi_x\|$. By some rearrangement of
terms in $H$, we may assume without loss of generality that
$\psi_x$ acts on $\mh_{\Lambda_0+x}$. Now, let
$A_x=h_x+\phi^{(r)}_x$, where
$$h_x=\|A\|P_{\mh_{\Lambda_0+x}\ominus\Omega_{\Lambda_0+x,0}},
\quad \phi^{(r)}_x=
A_x-\|A\|P_{\mh_{\Lambda_0+x}\ominus\Omega_{\Lambda_0+x,0}}.$$
Here and in the sequel $P_X$ stands for the projector onto $X$. It
follows that $\sum_xh_x$ is a classical Hamiltonian satisfying our
assumptions and, by the spectral gap condition on $A$,
$\sum_x\phi^{(r)}_x$ is its relatively bounded perturbation so
that (\ref{p2}) holds with $\alpha=(\|A\|-1)/\|A\|<1$. We consider
now $\psi_x$ as $\phi^{(b)}_x$, and then Theorem 2 applies.

Now we describe the application of Theorem 2 to the AKLT model.
This model was introduced by Affleck {\it et al.}
\cite{AKLT1,AKLT2} as the first rigorous example of a system in
the Haldane phase (\cite{H1,H2}, see \cite{Af} for a review of the
Haldane conjecture). It is a spin-1 chain with the
translation-invariant nearest-neighbor isotropic interaction
$$H=\sum_{k\in\zz}P^{(2)}({\bf S}_k+{\bf S}_{k+1})\equiv
\sum_{k\in\zz}({\bf S}_k\cdot{\bf S}_{k+1}/2+({\bf S}_k\cdot{\bf
S}_{k+1})^2/6+1/3),$$ where ${\bf S}_k$ is a spin-1 vector at site
$k$, and $P^{(2)}({\bf S}_k+{\bf S}_{k+1})$ is the projector onto
the subspace where ${\bf S}_k+{\bf S}_{k+1}$ has total spin 2. The
AKLT model has a unique gapped qround state $\omega$ minimizing
the energy of each term in the interaction: $$\omega(P^{(2)}({\bf
S}_k+{\bf S}_{k+1}))=0$$ (a frustration-free ground state). The
state $\omega$ can be described as a valence-bond-solid state
\cite{AKLT2} or a finitely correlated state \cite{FNW1,FNW2}. On a
finite chain $\Lambda$ with periodic boundary conditions the AKLT
Hamiltonian $H_\Lambda$ has a unique frustration-free ground
state.

Let $\Phi=\sum_k\phi_k$ be any translation-invariant finite range
interaction on the spin-1 chain. We consider the perturbed AKLT
model $H+\Phi$, starting, as before, with periodic finite chains
$\Lambda$. We prove

\begin{theor} If $\|\phi_k\|\le\beta$, with some $\beta$ depending
on the range of $\Phi$, then all conclusions of Theorem 1 hold for
the perturbed AKLT model $H+\Phi$.
\end{theor}

The main point of the proof is that at large scale the AKLT model
is a relatively bounded perturbation of a classical model. This
enables us to use Theorem 2. Though in this paper we restrict our
attention to the AKLT model only, this property is definitely more
general; one can expect some form of it to be generic to
non-critical gapped spin systems.

\section{Proof of Theorem 1}

We follow the standard approach and approximate the ground state
with low-temperature states. After time discretization we obtain a
cluster expansion, which identifies the model with a low density
hard-core gas of  excited regions on the space-time lattice
\cite{KT2}. After that all conclusions of Theorem 1 follow in a
usual way. Our exposition is, however, rather different
technically: we derive necessary cluster estimates using the
Schwarz lemma and resolvent expansions instead of the Feynman-Kac
formula.

We begin by proving that $\hl$ has a gapped ground state. Fix some
$t_0>0$ and consider the expectation
\begin{equation}\label{znl}Z_{N,\Lambda}\equiv\langle
(e^{-t_0H_\Lambda})^N\olo,\olo\rangle,\end{equation} at large
$N\in\mathbb N$. If $\ol$ is a non-degenerate ground state of
$\hl$ with the energy $E_\Lambda$ and a spectral gap $\ge \gamma$,
then $$Z_{N,\Lambda}=|\langle \ol,\olo\rangle|^2e^{-t_0E_\Lambda
N}+O(e^{-t_0(E_\Lambda+\gamma)N})$$ and hence, if $\langle
\ol,\olo\rangle\ne 0$, $$\ln Z_{N,\Lambda}=2\ln|\langle
\ol,\olo\rangle|-t_0E_\Lambda N+O(e^{-t_0\gamma N}).$$ Conversely,
if we show that for some constants $a_1,a_2,a_3$
\begin{equation}\label{as}\ln Z_{N,\Lambda}=a_1+a_2N+O(e^{-a_3N}),\end{equation}
with $a_3>0$, this will imply that in the cyclic subspace
generated by $\olo$ the operator $\hl$ has a gapped ground state.
We will argue later that the asymptotic (\ref{as}) holds, with the
same $a_2$ and $a_3$, if we add a small perturbation to $\olo$ in
(\ref{znl}), so $\hl$ has a gapped ground state in the whole space
$\mh_\Lambda$. The non-degeneracy of the ground state can be
deduced by a continuity argument from the non-degeneracy of the
ground state in the non-perturbed system.

We begin proving (\ref{as}) by writing the identity
$$e^{-t_0\hl}=\sum_{I\subset\Lambda%\{x:\Lambda_0+x\subset\Lambda\}
}T_{\Lambda,I},$$
where 
$$T_{\Lambda,I}=\sum_{J\subset
I}(-1)^{|I|-|J|}e^{-t_0(\hlo+\sum_{x\in
J}\phi_x)}.$$
Here the operator $H_{\Lambda,J}\equiv\hlo+\sum_{x\in J}\phi_x$ is
defined by the KLMN theorem, like $\hl$. When formally Trotter or
Duhamel expanded, $T_{\Lambda,I}$ is, by an inclusion-exclusion
argument, the contribution to the total evolution from the
perturbation of the classical evolution containing terms $\phi_x$
with $x\in I$ (see \cite{KT2}). We do not explicitly use these
expansions, however. Since all non-commutative terms $\phi_x$ in
$T_{\Lambda,I}$ lie in $\Lambda_I\equiv\cup_{x\in
I}(\Lambda_0+x)$, we can write
\begin{equation}\label{t'}
T_{\Lambda,I}=T'_Ie^{-t_0H_{\Lambda\setminus\Lambda_I,0}},\end{equation}
where $H_{\Lambda\setminus\Lambda_I,0}=
\sum_{x\in\Lambda:(\Lambda_0+x)\cap\Lambda_I=\varnothing}h_x$, and
$T'_I$ is defined as $T_{\Lambda,I}$ with $\hlo$ replaced by
$\sum_{x:(\Lambda_0+x)\cap\Lambda_I\ne\varnothing}h_x
=\hlo-H_{\Lambda\setminus\Lambda_I,0}$. For any
$\Lambda_1\subset\Lambda$ we will denote its neighborhood
$\cup_{x:(\Lambda_0+x)\cap\Lambda_1\ne\varnothing}(\Lambda_0+x)$
by $\widetilde\Lambda_1$ , so that $T'_I$ acts on
$\mh_{\widetilde\Lambda_I}$.

\begin{lemma}$\|T'_{I}\|\le(2\alpha
e^{t_0\beta/\alpha})^{|I|}.$
\end{lemma}
\begin{proof}
For some $J\subset I$, let $z_J\equiv
(z_{x_1},\ldots,z_{x_{|J|}}),x_k\in J,$ be a complex vector and
consider the operator-valued function
$$H_{J}(z_J)=\sum_{x:(\Lambda_0+x)\cap\Lambda_I\ne\varnothing}h_x
+\sum_{x\in J}z_x\phi_x.$$ If all $|z_x|<1/\alpha$, then, by
(\ref{phix}), the quadratic form $\sum_{x\in J}z_x\phi_x$ is
bounded relative to
$\sum_{x:(\Lambda_0+x)\cap\Lambda_I\ne\varnothing}h_x$, with a
relative bound $<1:$
\begin{equation}\label{zj}|\sum_{x\in J}z_x\phi_x(v,v)|\le
\max|z_x|\alpha \Bigl\|\Bigl(
\sum_{x:(\Lambda_0+x)\cap\Lambda_I\ne\varnothing}h_x\Bigr)^{1/2}v
\Bigr\|^2 +\max|z_x||J|\beta\|v\|^2.\end{equation}Therefore
$H_{J}(z_J)$ is an analytic family of m-sectorial operators on
$\mh_{\widetilde\Lambda_I}$ for $z_J\in\{|z_x|<1/\alpha|x\in J\}$
(see \cite{K}). Since by (\ref{zj}) the numerical range $\{\langle
H_{J}(z_J)v,v\rangle| v\in\dom(H_{J}(z_J)),\|v\|=1\}$ of these
operators lies in the half-plane $\{{\rm Re\,}z\ge
|J|\beta/\alpha\}$, it follows from the Hille-Yosida theorem that
\begin{equation}\label{et}
\|e^{-t_0H_{J}(z_J)}\|\le e^{t_0|J|\beta/\alpha}.\end{equation}
Now we consider the operator-valued function
$$T_{I}(z_I)=\sum_{J\subset I}(-1)^{|I|-|J|}e^{-t_0H_{J}(z_J)},$$
where $z_J$ is a restriction of $z_I$ to $J$. The function
$T_{I}(z_I)$ is analytic in $\{|z_x|<1/\alpha|x\in J\}$ and, by
(\ref{et}), $\|T_{I}(z_I)\|\le 2^{|I|}e^{t_0|I|\beta/\alpha}$.
Note that if $z_x=0$ for some $x\in I$, then $T_{I}(z_I)=0$
because in this case the terms $J\setminus\{x\}$ and $J\cup\{x\}$
make opposite contribution. Finally, $T'_{I}$ appearing in
(\ref{t'}) is the value of $T_{I}(z_I)$ at $z_I=(1,1,\ldots,1)$.
Now we use a many-dimensional version of the Schwarz lemma.

\begin{lemma} Let $f(z_I)$ be an analytic function in
$\{|z_x|<a|x\in I\}$ and $|f(z_I)|\le M$ for all $z_I$. Suppose
that if $z_x=0$ for some $x$, then $f(z_I)=0$. Then $|f(z_I)|\le
Ma^{-|I|}\prod_{x\in I}|z_x|$.
\end{lemma}

This lemma follows by induction from the usual one-dimensional
Schwarz lemma. Applying it to $T_{I}(z_I)$, we obtain the desired
estimate.
\end{proof}

By expanding $e^{-t_0\hl}$ in $T_{\Lambda,I}$ we have isolated the
regions with non-classical evolution; to obtain the final cluster
expansion we need to isolate in addition regions with classically
evolving excited states. Denote $\Lambda_I=\cup_{x\in
I}(\Lambda_0+x)$, and also  $$\mh_x'\equiv\mh_x\ominus\Omega_x,
\quad\mh'_{\Lambda_1}\equiv\otimes_{x\in\Lambda_1}\mh_x',$$ and
write $T_{\Lambda,I}$ as
$$T_{\Lambda,I}=T_{\Lambda,I}\sum_{J\subset\Lambda\setminus\Lambda_I}
P_{\mh_J'}P_{\Omega_{(\Lambda\setminus\Lambda_I)\setminus J,0}}.$$
Now define a configuration $C$ as a sequence
$\{(I_k,J_k)|k=1,\ldots,N\}$, where
$J_k\subset\Lambda\setminus\Lambda_{I_k}$; it follows that
$$Z_{N,\Lambda}=\sum_C w(C),$$ where
\begin{equation}\label{w}w(C)=\left\langle\prod_{k=1}^{N}
\left(T_{\Lambda,I_k} P_{\mh_{J_k}'}
P_{\Omega_{(\Lambda\setminus\Lambda_{I_k})\setminus
J_k,0}}\right)\olo,\olo\right\rangle\end{equation} with the
time-ordered product $\prod_{k=1}^{N}A_k\equiv A_{N}\cdots A_1$.

\begin{lemma}
$|w(C)|\le \prod_{k=1}^{N}\left((2\alpha
e^{t_0\beta/\alpha})^{|I_k|}
e^{-t_0(|J_k|-|\Lambda_0|^3|I_k|)}\right).$
\end{lemma}

\begin{proof}
 We estimate the norm of the operator in round brackets in (\ref{w}). By Lemma
 1,
$\|T'_{I}\|\le(2\alpha e^{t_0\beta/\alpha})^{|I|}$. Next, if
$J\subset\Lambda\setminus\Lambda_I$, then
$|J\cap(\Lambda\setminus\widetilde\Lambda_I)|
\ge|J|-|\Lambda_0|^2|\Lambda_I|\ge|J|-|\Lambda_0|^3|I|$. Any
$x\in\Lambda\setminus\widetilde\Lambda_I$ belongs to $|\Lambda_0|$
sets of the form $\Lambda_0+y$, these sets don't overlap with
$\Lambda_I$ and all contain $|\Lambda_0|$ sites; therefore by the
spectral gap assumption about $h_x$
$$H_{\Lambda\setminus\Lambda_I,0}\biggl|_
{\mh_J'\otimes{\Omega_{(\Lambda\setminus\Lambda_I)\setminus J,0}}}
\ge |J\cap(\Lambda\setminus\widetilde\Lambda_I)|{\bf
1}\ge(|J|-|\Lambda_0|^3|I|){\bf 1}.$$ It follows that the norm of
the expression in brackets in (\ref{w}) does not exceed $(2\alpha
e^{t_0\beta/\alpha})^{|I_k|} e^{-t_0(|J_k|-|\Lambda_0|^3|I_k|)}$,
which implies the desired estimate.
\end{proof}

Now for a configuration $C$ we define its support $\supp
C\subset\{0,1,\ldots,N\}\times\Lambda$ as the set
\begin{equation}\label{supp}\{(k,x)|k=0,\ldots,N;x\in\widetilde\Lambda_{I_{k}}
\cup\widetilde\Lambda_{I_{k+1}}\cup\widetilde J_{k}\cup\widetilde
J_{k+1}\}\end{equation} (with $\widetilde\Lambda_{I_{0}}
=\widetilde\Lambda_{I_{N+1}}=\widetilde J_{0}=\widetilde
J_{N+1}=\varnothing$). We say that configurations are disjoint if
they have disjoint supports. If $C_1$ and $C_2$ are disjoint, we
naturally define their union $C=C_1\cup C_2$ as the configuration
with $I_k=I_k^{(1)}\cup I_k^{(2)},J_k=I_k^{(1)}\cup J_k^{(2)}.$

\begin{lemma}
If $C_1$ and $C_2$ are disjoint, then $w(C_1\cup
C_2)=w(C_1)w(C_2)$.
\end{lemma}

\begin{proof} For the configuration $C=C_1\cup C_2$ and any $n=1,\ldots,N$
consider the vector $$v_n=\prod_{k=1}^{n} \left(T_{\Lambda,I_k}
P_{\mh_{J_k}'} P_{\Omega_{(\Lambda\setminus\Lambda_{I_k})\setminus
J_k,0}}\right)\olo,$$ so that $w(C)=\langle v_{N},\olo\rangle$. We
have $v_n=u_n\otimes\Omega_{\Lambda\setminus(\Lambda_{I_n}\cup
J_n),0}$ with some $u_n\in\mh_{\Lambda_{I_n}\cup J_n}$.
Analogously, we can define
$v^{(1)}_n,v^{(2)}_n,u^{(1)}_n,u^{(2)}_n$ for $C_1,C_2$. Let
$K_n=\Lambda_{I_n}\cup J_n$ and similarly define
$K_n^{(1)},K_n^{(2)}$ for $C_1,C_2$. Since $\supp C_1$ and $\supp
C_2$ are disjoint, it follows in particular that $K_n^{(1)}$ and
$K_n^{(2)}$ are disjoint, so that $\mh_{K_n}= \mh_{K_n^{(1)}}
\otimes \mh_{K_n^{(2)}}$. We will prove by induction that
$u_n=u_n^{(1)}\otimes u_n^{(2)}$; at $n=N$ this implies the
desired equality $w(C_1\cup C_2)=w(C_1)w(C_2)$. Suppose that
$u_{n-1}=u_{n-1}^{(1)}\otimes u_{n-1}^{(2)}$. Note that we have in
$\mh_{\widetilde K_n\cup K_{n-1}}$ the equality
\begin{equation}\label{un}
u_n\otimes\Omega_{(\widetilde K_n\cup K_{n-1})\setminus
K_n,0}=T_{K_n}'P_{\Omega_{(\widetilde K_n\cup K_{n-1})\setminus
K_n,0}}(u_{n-1}\otimes\Omega_{(\widetilde K_n\cup
K_{n-1})\setminus K_{n-1},0}),\end{equation} where $T_{K_n}'$ is
defined as $T_{\Lambda,I}$ with $\hlo$ replaced by
$\sum_{x:(\Lambda_0+x)\cap K_n\ne\varnothing}h_x
=\hlo-H_{\Lambda\setminus K_n,0}$. (\ref{un}) holds because
$e^{-t_0(\hlo-H_{\Lambda\setminus K_n,0})}$ acts trivially on the
ground state. By the disjointness, the objects in (\ref{un})
factor into products of respective objects for $C_1,C_2$, which
proves the inductive step.
\end{proof}

A polymer $\chi$ is a connected configuration (i.e., which is not
a union of two configurations with disjoint supports). We have $$
Z_{N,\Lambda}=\sum_{\text{disj.
}\chi_1,\ldots\chi_n}\nolimits^{N,\Lambda}\prod_{k=1}^n
w(\chi_n),$$ where summation is over all disjoint collections of
polymers in $\{1,\ldots,N\}\times\Lambda$. This is the desired
polymer expansion. By Lemma 3, for any $\epsilon>0$ we can choose
$t_0$ large and then $\alpha,\beta$ small so that
$w(\chi)\le\epsilon^{|\supp\chi|}$. A standard combinatorial
argument shows that the number of polymers with $|\supp \chi|=n$
containing a given point does not exceed $c^n$ for some
$c=c(\nu,\Lambda_0)$. Now all conclusions of Theorem 1 follow from
standard results on cluster expansions \cite{KP,MM,Sei,Sim,KT2},
and we will be very sketchy. We define a cluster $X$ as a
connected collection of polymers $\chi_1,\ldots,\chi_k$ with
positive multiplicities $n_1,\ldots,n_k$. Let $G(X)$ be a graph
with $n_1+\ldots+n_k$ vertices, corresponding to these polymers,
and a line between two vertices drawn if the corresponding
polymers intersect. Let $G_1\vartriangleleft G(X)$ stand for a
connected subgraph $G_1$ containing all vertices of $G(X)$, and
$l(G_1)$ be the number of lines in $G_1$. Then the weight of the
cluster $X$ is defined as $$w(X)=(n_1!\cdots
n_k!)^{-1}w(\chi_1)^{n_1}\cdots
w(\chi_k)^{n_k}\sum_{G_1\vartriangleleft G(X)}(-1)^{l(G_1)}.$$ It
follows that $$\ln \sum_{\text{disj.
}\chi_1,\ldots\chi_n}\nolimits^{N,\Lambda}\prod_{k=1}^n
w(\chi_n)=\sum_X\nolimits^{N,\Lambda}w(X),$$ with the absolutely
convergent series on the r.h.s. (see \cite{BZ,U} for recent simple
proofs). Let $l(X)$ be the time length of a cluster; shifting
clusters in time, we write
$$\sum_X\nolimits^{N,\Lambda}w(X)=\sum_{X:l(X)\le
N}\nolimits^{t=0,\Lambda} (N-l(X))w(X)$$ $$
=-\sum_{X}\nolimits^{t=0,\Lambda}l(X)w(X)+
N\sum_{X}\nolimits^{t=0,\Lambda}w(X)+\sum_{X:l(X)>
N}\nolimits^{t=0,\Lambda}(l(X)-N)w(X),$$ where
$\sum^{t=0,\Lambda}$ is the sum over clusters starting at $t=0$.
By the cluster estimate, all series in the r.h.s. converge
absolutely, and the last term is $O(\epsilon^N)$ because summation
is over clusters with length $>N$. Comparing this with (\ref{as}),
we identify $a_1$ as $-\sum_{X}\nolimits^{t=0,\Lambda}l(X)w(X)$,
$a_2$ as $\sum_{X}\nolimits^{t=0,\Lambda}w(X)$, and $a_3$ as
$-\ln\epsilon$. This $\epsilon$ does not depend on $\Lambda$, so
the spectral gap estimate is volume-independent. To complete the
proof of 1), we consider the changes in the asymptotic (\ref{as})
when $\olo$ is replaced by $\olo+v$ with small $v$. This
replacement adds new polymers $\chi_v$, arising from the new terms
$ \langle (e^{-t_0H_\Lambda})^Nv,\olo\rangle$, $ \langle
(e^{-t_0H_\Lambda})^N\olo,v\rangle$ and $ \langle
(e^{-t_0H_\Lambda})^Nv,v\rangle$. The support of $\chi_v$ contains
$\{0\}\times\Lambda$ or $\{N\}\times\Lambda$, or both. For $v$
small enough the estimate $|w(\chi)|\le \epsilon^{|\supp\chi|}$
remains valid for $\chi_v$. The expansion for $\ln Z_{N,\Lambda}$
is modified by adding clusters containing the new polymers
$\chi_v$. Such clusters touch the boundary of the time segment
$\{0,\ldots,N\}$, and hence their contribution is
$c+O(\epsilon^N)$. This completes the proof of 1).
To show that 2) holds, one writes $$\langle
A\ol,\ol\rangle=\lim_{N\to\infty}Z_{2N,\Lambda}^{-1}\langle A
e^{-t_0 N H_\Lambda}\olo,e^{-t_0 N H_\Lambda}\olo\rangle,$$ using
the fact that $\langle\olo,\ol\rangle\ne 0$. If $A$ acts on
$\mh_{\Lambda_A}$, one introduces new polymers $\chi_A$ with the
support containing $\{0\}\times\Lambda_A$ and the weight
calculated using $A$ inserted in the 0th layer; one has
$|w(\chi_A)|\le\epsilon^{|\supp\chi_A|-|\Lambda_A|}\|A\|$. It
follows that $\langle
A\ol,\ol\rangle=\sum_X\nolimits^{\Lambda}w(X)$, where the sum is
over clusters in $\zz\times\Lambda$, containing one polymer
$\chi_A$ with multiplicity 1. As $\Lambda\nearrow\znu$, this
expression tends to the absolutely convergent sum over polymers in
$\zz\times\znu$, which proves 2).  3) follows from the fact that
the truncated correlation on the l.h.s. equals $\sum_Xw(X)$ over
clusters containing either a polymer with the support containing
$\{0\}\times\Lambda_1$ and $\{0\}\times\Lambda_2$, or two polymers
$\chi_{A_1},\chi_{A_2}$. Finally, 4) follows because if $\phi$
varies analytically, then the cluster expansion does too and is
convergent as long as the estimate (\ref{phix}) holds.

\section{Proof of Theorem 2}
Following \cite{KT2}, in order to extend the perturbation theory
to $\alpha$ close to 1 we use a scaling transformation. We group
lattice sites in cubic blocks $b_x$ of linear size $l$, so that
the initial cubic volume $\Lambda$ is transformed into cubic
volume $\ovl$ whose sites $x$ are these blocks (we assume that the
size of $\Lambda$ is a multiple of $l$, but one can consider
general cubic volumes too by taking blocks of different sizes).
For any $x\in\ovl$, let $$\ovh_x=\otimes_{y\in
b_x}\mh_y,\quad\ovo_x=\otimes_{y\in
b_x}\Omega_y,\quad\ovh_x'=\ovh_x\ominus\ovo_x$$ and also
$$\ovh_I'=\otimes_{x\in
I}\ovh_x',\quad\ov\Omega_{I,0}=\otimes_{x\in I}\ov\Omega_x$$ for
$I\subset\ovl$. Suppose that $l>\diam(\Lambda_0)$. We can then
view the interaction $\Phi_\Lambda$ as the sum
$\sum_{x\in\ovl}\ov{\phi}_x$; here $\ov{\phi}_x$ acts on
$\ov\mh_{\ovl_0+x}$, where $\ovl_0=\{0,1\}^\nu$, and is defined as
\begin{equation}\label{op}\ov{\phi}_x=
\sum_{y:\Lambda_0+y\in\cup_{z\in\ovl_0+x}b_z}c_y^{-1}\phi_y.\end{equation}
Here $c_y=|\{x:\Lambda_0+y\in\cup_{z\in\ovl_0+x}b_z\}|$ is the
number of $\ov\phi_x$ where $\phi_y$ appears. We define
analogously $\ov{\phi}_x^{(r)},\ov{\phi}_x^{(b)}$ for
${\phi}_x^{(r)},{\phi}_x^{(b)}$ appearing in the decomposition
(\ref{p1}). Next for any $I\subset\ovl$ we define
$$\ov{\Phi}_{I}=\sum_{x\in I}\ov{\phi}_x$$ and similarly for
$\ov{\Phi}_{I}^{(r)},\ov{\Phi}_{I}^{(b)}$, so that
$H_\Lambda=\hlo+\ov{\Phi}_{\ovl}^{(r)}+\ov{\Phi}_{\ovl}^{(b)}.$
Following the proof of Theorem 1, we write
$$e^{-t_0H_\Lambda}=\sum_{I,J,K\subset\ovl}T_{I,J,K},$$ where
$$T_{I,J,K}=\sum_{I_1\subset I}\sum_{J_1\subset
J}(-1)^{|I|-|I_1|}(-1)^{|J|-|J_1|}e^{-t_0(\hlo+\opi+\opj)}
P_{\ovh_K'\otimes\ovo_{\ovl\setminus K,0}}.$$ We call a sequence
$C=\{(I_k,J_k,K_k),k=1,\ldots, N\}$ a configuration and assign to
it the weight
$w(C)=\langle\prod_{k=1}^N\ov{T}_{I_k,J_k,K_k}\ovo_{\ovl,0},\ovo_{\ovl,0}\rangle$.
Let $x_1,\ldots,x_\nu$ stand for the coordinates of the site
$x\in\ovl$; for any set $I\subset\ovl$ we define its neighborhood
$$\widetilde{I}=\{x|\exists y\in I\text{ such that } |x_k-y_k|\le
1, k=1,\ldots,\nu\}.$$ Let $\ovl_I=\cup_{x\in I}(\ovl_0+x)$.
Similarly to (\ref{supp}), we define a configuration's support as
$$\{(k,x)|k=0,\ldots,N;x\in\widetilde\ovl_{I_{k}}
\cup\widetilde\ovl_{I_{k+1}}\cup\widetilde\ovl_{J_{k}}
\cup\widetilde\ovl_{J_{k+1}}\cup\widetilde{K}_{k}
\cup\widetilde{K}_{k+1}\}.$$ An analog of Lemma 4 on factorization
of  weights is immediate. Therefore the only thing that needs to
be proved is an exponential bound for the weight:
$$|w(C)|\le\epsilon^{|\supp C|}$$ with sufficiently small
$\epsilon$; after that the conclusion of the theorem follows like
in the previous section. Clearly, in order to have this bound it
suffices to show that for any $\epsilon$ one can choose $t_0,l$
and $\beta$ so that
\begin{equation}\label{tijk}
\|T_{I,J,K}\|\le\epsilon^{|I|+|J|+|K|}.\end{equation} The
remainder of this section is a proof of this claim. Specifically,
we will show that for $\alpha$ close to 1 one can achieve this by
choosing
\begin{equation}\label{tlb}t_0=(1-\alpha)^{-\varkappa},\quad
l=\lceil(1-\alpha)^{-\varkappa}\rceil,\quad
\beta=\delta(1-\alpha)^{\varkappa(\nu+1)},\end{equation} where
 $\lceil\cdot\rceil$ is
the integer part; here $\varkappa$ is any fixed constant $>1$, and
$\delta=\delta(\varkappa,\nu,\Lambda_0)$ a sufficiently small
constant. The strategy of the proof is as follows. We will obtain
three different bounds for $\|T_{I,J,K}\|$, suitable when the
contribution to $I\cup J\cup K$ from either $I,J$ or $K$ is large
enough.

{\bf Case 1.} The first bound relies on the smallness of the
bounded part $\ov{\phi}^{(b)}$ of the perturbation, and is used
when $J$ is large. In this case we use again the Schwarz lemma. In
the definition of $T_{I,J,K}$ we replace $\sum_{x\in
J_1}\ov{\phi}_x^{(b)}$ with $\sum_{x\in J_1}z_x\ov{\phi}_x^{(b)}$,
where $z_x\in\cc, |z_x|\le a$, with some $a>1$. Let
$a=(t_0\|\ov{\phi}^{(b)}\|)^{-1}$, then by definition of
$\ov{\phi}^{(b)}$ and from (\ref{tlb})
$$a\ge(t_0(2l)^\nu\|\phi^{(b)}\|)^{-1}\ge(t_0(2l)^\nu\beta)^{-1}\ge
2^{-\nu}\delta^{-1}> 1$$ if $\delta<2^{-\nu}$. Using the Schwarz
lemma with this $a$, we find that
\begin{equation}\label{case1}\|T_{I,J,K}\|\le
2^{|I|}(2et_0\|\ov{\phi}^{(b)}\|)^{|J|}\le
2^{|I|}(2^{\nu+1}e\delta)^{|J|}.\end{equation}

{\bf Case 2.} The second bound relies on the contractiveness of
the classical evolution in excited regions and is used when $K$ is
large. We will estimate the norm of $e^{-t_0(\hlo+\opi+\opj)}
P_{\ovh_K'\otimes\ovo_{\ovl\setminus K,0}}.$ We begin by writing
\begin{equation}\label{eint}
e^{-t_0(\hlo+\opi+\opj)}=(2\pi i)^{-1}\int_\Gamma
e^{-t_0z}R_zdz,\end{equation} where $R_z$ is the resolvent
$(\hlo+\opi+\opj-z)^{-1}$, and $\Gamma$ is a contour in the
complex plane going around the spectrum of $\hlo+\opi+\opj$; we
will specify $\Gamma$ below. We will use the expansion of the
resolvent
\begin{equation}\label{res} R_z=Q_z(\sum_{k=0}^\infty
F_z^k)Q_z,\end{equation} where
$$F_z=-(\hlo+\opj-z)^{-1/2}\opi(\hlo+\opj-z)^{-1/2}$$ and
$$Q_z=(\hlo+\opj-z)^{-1/2}.$$ The operators $Q_z$ are well-defined
and bounded for $z$ in the resolvent set of $\hlo+\opj$, i.e., at
least in $\cc\setminus [-\|\opj\|,+\infty)$. We now estimate the
norm of $F_z$:

\newpage
\begin{eqnarray*}\|F_z\| & = &
\||\hlo+\opj-z|^{-1/2}\opi|\hlo+\opj-z|^{-1/2}\|\\ & = &
\sup_{u\in\ovh_{\ovl}\setminus\{0\}}
\frac{|\langle|\hlo+\opj-z|^{-1/2}\opi|\hlo+\opj-z|^{-1/2}u,u\rangle|}
{\|u\|^2}\\ & = & \sup_{v\in\dom(|\hlo+\opj-z|^{1/2})}
\frac{|\opi(v,v)|}{\||\hlo+\opj-z|^{1/2}v\|^2}\\ & \le &
\sup_{v\in\dom(|\hlo+\opj-z|^{1/2})}
\frac{\alpha\|\hlo^{1/2}v\|^2}{\||\hlo+\opj-z|^{1/2}v\|^2}\\ & \le
&  \sup_{\lambda\in\spec(\hlo+\opj)}
\frac{\alpha(\lambda+\|\opj\|)}{|\lambda-z|}.
\end{eqnarray*} We
will be interested in those $z$ where $\|F_z\|\le\sqrt{\alpha}$.
By the above bound, a sufficient condition for that is
\begin{equation}\label{u}z\notin \bigcup_{\lambda\in\spec(\hlo+\opj)}
\Bigl\{z\in\cc\Bigl||z-\lambda|\le\sqrt{\alpha}(\lambda+\|\opj\|)\Bigr\}.
\end{equation} Since $\spec(\hlo+\opj)\subset[-\|\opj\|,+\infty)$, the above
union lies in the sector
\begin{equation}\label{sec}\{z\in\cc: |\arg
(z+\|\opj\|)|\le\arcsin\sqrt{\alpha}\},\end{equation} so if we
choose $z$ outside this sector we have $\|F_z\|\le\sqrt{\alpha}$
and in particular
\begin{equation}\label{fk0} \|Q_zF_z^kQ_z\|
\le\frac{\alpha^{k/2}}{\dist(z,\spec(\hlo+\opj))}.\end{equation}

We will now show that, furthermore, one can enlarge the domain of
$z$ where a bound of the above type holds, if one applies the
operator on the l.h.s. to vectors from
$\ovh_K'\otimes\ovo_{\ovl\setminus K,0}$ with $K$ large compared
to $J$. Precisely, let $$n=\lceil (|K|-6^\nu|J|)/7^\nu\rceil.$$
This $n$ is a lower bound for the maximal number of sites in a
subset $K_1\subset K$ such that the neighborhoods
$\widetilde{\{x\}}$ of points $x\in K_1$ are separated from each
other and from $\ovl_J$ by at least two $\ovl$-lattice spacings:
choose the first such $x$ outside the 2-neighborhood of $\ovl_J$,
then the second outside the 2-neighborhood of $\ovl_J$ unioned
with the 3-neighborhood of the first $x$, etc. We assume that
$n>0$. Next, let $$m=\lceil l/\diam(\Lambda_0)\rceil,m>0.$$ For
any $a$ define $\mathcal U_a$ as the union of circles standing in
(\ref{u}), but with $\lambda$ running over $[a,+\infty)$. Now,
suppose that $\lceil k/m\rceil=r$ with $r\le n$; we claim then
that
\begin{eqnarray}\label{fk}\|Q_zF_z^kQ_z
P_{\ovh_K'\otimes\ovo_{\ovl\setminus K,0}}\|
\le\frac{\alpha^{k/2}}{\dist(z,[|\Lambda_0|(n-r)-\|\opj\|,+\infty))},
\\\quad \forall z\notin\mathcal
U_{|\Lambda_0|(n-r)-\|\opj\|}.\nonumber \end{eqnarray}

Indeed, let $\mathcal G_{[a,+\infty)}$ stand for the spectral
subspace of $\hlo+\opj$, corresponding to the interval
$[a,+\infty)$. Let $K_1$ be chosen as above, $|K_1|=n$. Then the
subspace $\ovh_{{K}_1}'\otimes \ovh_{\ovl\setminus{K}_1}$ is an
invariant subspace of $\hlo+\opj$ and it lies in $\mathcal
G_{[|\Lambda_0|n-\|\opj\|,+\infty)}$. The sites $x\in K_1$ (and
therefore their neighborhoods $\widetilde{\{x\}}$ too) are in
excited states. When we apply powers of $F_z$ to vectors from
$\ovh_{{K}_1}'\otimes \ovh_{\ovl\setminus{K}_1}$, we need at least
$m=\lceil l/\diam(\Lambda_0)\rceil$ powers to remove the
excitation from  a neighborhood $\widetilde{\{x\}}$ of a given
$x\in K_1$, because the block $b_x$ has to be connected to
$\Lambda\setminus\cup_{y\in\widetilde{\{x\}}} b_y$ by supports of
the elementary interactions $\phi^{(r)}_w,w\in\Lambda,$ of which
$\opi$ is composed. Hence to remove excitations from $r$
neighborhoods we need at least $mr$ powers of $F_z$. Therefore if
$k<mr$ with $r\le n$, then $$F_z^k
{\ovh_K'\otimes\ovo_{\ovl\setminus K,0}}\subset \mathcal
G_{[|\Lambda_0|(n-r)-\|\opj\|,+\infty)}$$ (here we use
$\ovh_K'\otimes\ovo_{\ovl\setminus K,0}\subset
\ovh_{{K}_1}'\otimes \ovh_{\ovl\setminus{K}_1}$). It follows that
in the case at hand we can replace $\hlo+\opj$ and the quadratic
form $\opi$ with their restrictions to $\mathcal
G_{[|\Lambda_0|(n-r)-\|\opj\|,+\infty)}$; the argument leading to
(\ref{fk0}) then yields (\ref{fk}).

Now we can specify the integration contour $\Gamma$. It will
depend on the term $k$ in the expansion (\ref{res}) through
$r=\lceil k/m\rceil$. Let $s=n-r$. For $s=0,\ldots,n$, let
$$z_{s,\pm}=(1-\sqrt{\alpha})|\Lambda_0|s-\|\opj\|\pm
i\frac{\sqrt{\alpha}}{\sqrt{1-\alpha}}|\Lambda_0|s-t_0^{-1}$$ and
let the contour $\Gamma_s$ consist of the segment
$[z_{s,-},z_{s,+}]$ and the two rays
$\{z|\arg(z-z_{s,\pm})=\pm\arcsin\sqrt{\alpha}\}$, so that it is
the boundary of the truncated sector (\ref{sec}) (of the sector
itself in the case $s=0$) shifted by $t_0^{-1}$ to the left. This
contour is just a convenient for calculations, piecewise-linear
approximation of the boundary of $\mathcal
U_{|\Lambda_0|s-\|\opj\|}$. We choose the contour in this way
because we need it to lie as far to the right as possible due to
the factor $e^{-t_0z}$ in (\ref{eint}), but still outside of
$\mathcal U_{|\Lambda_0|s-\|\opj\|}$, so that the resolvent
estimate can be used. By slightly shifting it to the left we avoid
the possible singularity in the denominator in (\ref{fk}).

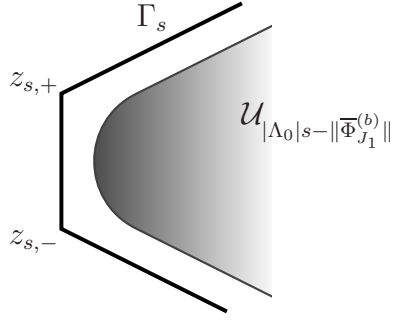
\begin{pspicture}(8,5.2)

\put(2,-0.5){\psline[linewidth=1.5pt]{-}(4,1)(1.8,2.1)(1.8,3.9)(4.2,5.1)}

\put(2,-0.5){\psline[linearc=1,linecolor=darkgray,fillstyle=gradient,
gradangle=90,gradbegin=darkgray,gradend=white](5,1)(1,3)(5,5)}

\put(2,-0.5){\pspolygon*[linecolor=white](5.2,0.7)(4.6,1.0)(4.6,5)(5.2,5.2)}

\put(4.8,4.3){$\Gamma_s$}

\put(6.2,3){$\mathcal U_{|\Lambda_0|s-\|\opj\|}$}

\put(3.1,1.4){$z_{s,-}$} \put(3.1,3.5){$z_{s,+}$}

\end{pspicture}

\noindent We have $\Gamma_s\cap\mathcal
U_{|\Lambda_0|s-\|\opj\|}=\varnothing$, so we can write
$$e^{-t_0(\hlo+\opi+\opj)}P_{\ovh_K'\otimes\ovo_{\ovl\setminus
K,0}}$$ $$=(2\pi i)^{-1}\sum_{r=0}^{n-1}\int_{\Gamma_{n-r}}
e^{-t_0z}Q_z\biggl(\sum_{k=rm}^{(r+1)m-1}
F_z^k\biggr)Q_zP_{\ovh_K'\otimes\ovo_{\ovl\setminus K,0}} dz$$

$$+(2\pi i)^{-1}\int_{\Gamma_0} e^{-t_0z}Q_z\biggl(\sum_{
k=nm}^\infty F_z^k\biggr)Q_zP_{\ovh_K'\otimes\ovo_{\ovl\setminus
K,0}} dz.$$ We estimate now this expression using $|\int e^{-t_0
z}f(z)dz|\le\int e^{-t_0{\,\rm Re\, }z}|f(z)||dz|$ and the bound
(\ref{fk}). We have
$\dist(\Gamma_s,[|\Lambda_0|s-\|\opj\|,+\infty))\ge\sqrt{\alpha}$.
It follows that
$$\|e^{-t_0(\hlo+\opi+\opj)}P_{\ovh_K'\otimes\ovo_{\ovl\setminus
K,0}}\|$$
$$\le\frac{e^{t_0\|\opj\|+1}}{2\pi\sqrt{1-\alpha}(1-\sqrt{\alpha})}
\Bigl(\sum_{r=0}^{n-1}\bigl[(\alpha^{\frac{rm}{2}}-\alpha^{\frac{(r+1)m}{2}})
e^{-t_0(1-\sqrt{\alpha})|\Lambda_0|(n-r)} \bigl(|\Lambda_0|(n-r)$$
$$ +\frac{2}{\sqrt{\alpha}t_0}\bigr)\bigl]
+\frac{2\alpha^{\frac{nm}{2}}}{\sqrt{\alpha}t_0}\Bigr) $$
$$\le\frac{e^{t_0\|\opj\|+1}}{2\pi\sqrt{1-\alpha}}
(n+1)\bigl(|\Lambda_0|n+\frac{2}{\sqrt{\alpha}t_0}\bigr)\bigl(
\max\{\alpha^{\frac{m}{2}},e^{-t_0(1-\sqrt{\alpha})|\Lambda_0|}\}\bigr)^{n}.$$
Some calculation now shows that if $t_0,l,\beta$ are defined as in
(\ref{tlb}) with $\varkappa>1$,  then for any $\epsilon>0$ we have
$$\|e^{-t_0(\hlo+\opi+\opj)}P_{\ovh_K'\otimes\ovo_{\ovl\setminus
K,0}}\|\le e^{t_0\|\opj\|}\epsilon^n\le
e^{2^\nu\delta|J_1|}\epsilon^n$$ when $\alpha>\alpha_0$ for some
$\alpha_0=\alpha_0(\epsilon,\varkappa,\nu,\Lambda_0)<1$. It
follows that \begin{equation}\label{case2}\|T_{I,J,K}\|\le
2^{|I|+|J|}e^{2^\nu\delta|J|}\epsilon^n.\end{equation}

{\bf Case 3.} The third bound is used when $I$ is large. We will
bound $\|\sum_{I_1\subset
I}(-1)^{|I|-|I_1|}e^{-t_0(\hlo+\opi+\opj)}
P_{\ovh_K'\otimes\ovo_{\ovl\setminus K,0}}\|.$ Like in case 2, we
represent $$e^{-t_0(\hlo+\opi+\opj)}=(2\pi i)^{-1}\int_{\Gamma_0}
e^{-t_0z}R_{z,I_1}dz,$$ where $R_{z,I_1}=(\hlo+\opi+\opj-z)^{-1}$
and $\Gamma_0$ is the shifted boundary of the sector defined in
case 2. We again expand $R_{z,I_1}=Q_z(\sum_{k=0}^\infty
F_{z,I_1}^k)Q_z$. Now, let $$n'=\lceil
(|I|-6^\nu|J|-5^\nu|K|)/6^\nu\rceil.$$ This $n'$ is a lower bound
for the maximal number of sites in a subset $I_2\subset I$ such
that the neighborhoods $\widetilde{\{\ovl_0+x\}}$ of points $x\in
I_2$ are separated from each other, from $\ovl_J$ and from $K$.
Let $m=\lceil l/\diam(\Lambda_0)\rceil$ as before. We claim that
$$\sum_{I_1\subset I}(-1)^{|I|-|I_1|}F_{z,I_1}^kQ_z
P_{\ovh_K'\otimes\ovo_{\ovl\setminus K,0}}=0$$ if $k<mn'$. Indeed,
if in the above sum we expand $F_{z,I_1}=-Q_z\sum_{x\in
I_1}\ov{\phi}_x^{(r)}Q_z$, then  by the inclusion-exclusion
principle it becomes
$$(-1)^k\sum\bigl(\prod_{p=1}^k(Q_z\ov{\phi}_{x_p}^{(r)}Q_z)\bigr)Q_z
P_{\ovh_K'\otimes\ovo_{\ovl\setminus K,0}},$$ where summation is
over sequences $x_1,\ldots,x_k$ of points from $I$, containing
each point of $I$ at least once. In particular, each sequence
contains all points of $I_2$ defined above. Each term in this sum
is 0. Indeed, if we further expand each $\ov{\phi}_x^{(r)}$ in
$\phi^{(r)}_y$ as in (\ref{op}) and consider the set
$\cup_{p=1}^k(\Lambda_0+y_p)$ for each of the resulting sequences,
then, if $k<mn'$, this set will have at least one connected
component contained in the set $\widetilde{\{\ovl_0+x\}}$ for some
$x\in I_2$. By the choice of $I_2$, the operator $\phi^{(r)}_y$
from this component which acts first acts on the ground state.
Since ${\phi}_y^{(r)}\Omega_{\Lambda_0+y}=0$, this implies our
claim.

It follows that $$\Bigl\|\sum_{I_1\subset
I}(-1)^{|I|-|I_1|}e^{-t_0(\hlo+\opi+\opj)}
P_{\ovh_K'\otimes\ovo_{\ovl\setminus K,0}}\Bigr\|$$
$$=\Bigl\|\sum_{I_1\subset I}(-1)^{|I|-|I_1|} (2\pi
i)^{-1}\int_{\Gamma_0}e^{-t_0z}Q_z\sum_{k=mn'}^{\infty}F_{z,I_1}^kQ_z
P_{\ovh_K'\otimes\ovo_{\ovl\setminus K,0}}dz\Bigr\|$$
$$\le\frac{2^{|I|}e^{t_0\|\opj\|+1}\alpha^{\frac{mn'}{2}}}{\pi
t_0\sqrt{\alpha(1-\alpha)}(1-\sqrt{\alpha})}.$$ Again, for any
$\epsilon>0$, if $\alpha$ is sufficiently close to 1, then this
expression does not exceed
$2^{|I|}e^{2^\nu\delta|J_1|}\epsilon^{n'}$, which implies
\begin{equation}\label{case3}\|T_{I,J,K}\|\le
2^{|I|+|J|}e^{2^\nu\delta|J|}\epsilon^{n'}\end{equation} in case
3.

The desired bound (\ref{tijk}) now follows from bounds obtained in
the three cases. If $I$ is sufficiently large compared to $J$ and
$K$, then one uses the bound (\ref{case3}). Otherwise, and if $K$
is sufficiently large compared to $J$, one uses (\ref{case2}). In
the remaining case one uses (\ref{case1}).

\section{Proof of Theorem 3}

We summarize some known facts about the AKLT model which we will
need. Denote by $H_\Lambda^{p}$ and $H_\Lambda^f$  the AKLT
Hamiltonians on a finite chain $\Lambda$ with periodic and free
boundary conditions, respectively. The Hamiltonian $H_\Lambda^f$
has a four-dimensional subspace $\mathcal G_\Lambda$ of
frustration-free ground states. Using the valence-bond-solid
representation, one can choose a (non-orthogonal) basis
$\Omega_{\Lambda;ab},a,b=1,2,$ in $\mathcal G_\Lambda$ with the
following properties:

1) For two adjacent finite chains $\Lambda_1,\Lambda_2$
$$\Omega_{\Lambda_1\cup\Lambda_2;ab}=
\Omega_{\Lambda_1;a1}\otimes\Omega_{\Lambda_2;2b} -
\Omega_{\Lambda_1;a2}\otimes\Omega_{\Lambda_2;1b}.$$  The unique
ground state of $H_\Lambda^{p}$ is given by
$\Omega_{\Lambda;12}-\Omega_{\Lambda;21}$.

2) Let $g_\Lambda$ be the $4\times 4$ Gram matrix of the basis
$\Omega_{\Lambda;ab}$: $$(g_\Lambda)_{ab,cd}
=\langle\Omega_{\Lambda;ab},\Omega_{\Lambda;cd}\rangle.$$ Then
$g_\Lambda={\bf 1}+O(3^{-|\Lambda|})$ as $|\Lambda|\to\infty$.

We will also use the fact that the operators $H_\Lambda^f$ have a
uniformly bounded away from 0 spectral gap, i.e. for some
$\gamma>0$ we have $H_\Lambda^f\ge\gamma ({\bf 1}-G_\Lambda)$ for
all $\Lambda$, where $G_\Lambda$ is the projector onto $\mathcal
G_\Lambda$.

All the above facts were proved in \cite{AKLT2}, see also
\cite{AAH,dNR,FNW2,KLT,KT2,Kn,N} for various refinements.

We show now that at large scale the AKLT model is a perturbation
of a non-interacting model. Like in the previous section, we group
the sites of the cyclic chain $\Lambda$ in blocks
$\Lambda_1,\Lambda_2,\ldots,\Lambda_n$ with
$|\Lambda_k|=l,k=1,\ldots,n$.  We will specify $l$ later. We write
$$H_\Lambda^p=\sum_{k=1}^n\overline{H}_{k,k+1},$$ where
$$\overline{H}_{k,k+1}=H_{\Lambda_k}^f/2+P^{(2)}({\bf S}_{kl}+{\bf
S}_{kl+1})+H_{\Lambda_{k+1}}^f/2$$
$$=H_{\Lambda_k\cup\Lambda_{k+1}}^f/2+P^{(2)}({\bf S}_{kl}+{\bf
S}_{kl+1})/2$$ (with the convention $n+1\equiv 1$). Clearly,
$\kerr(\overline{H}_{k,k+1})=
\kerr(H_{\Lambda_k\cup\Lambda_{k+1}}^f) =\mathcal
G_{\Lambda_k\cup\Lambda_{k+1}}$ and
\begin{equation}\label{sg}\overline{H}_{k,k+1}\ge
H_{\Lambda_k\cup\Lambda_{k+1}}^f/2\ge\gamma/2({\bf
1}-G_{\Lambda_k\cup\Lambda_{k+1}}).\end{equation} Now, an
important role is played by the asymptotic commutativity of the
projectors $G_{\Lambda_k\cup\Lambda_{k+1}}$, which we utilize as
follows. For each $k$ we orthogonalize the basis
$\Omega_{\Lambda_k;ab}$:
$$\Omega'_{\Lambda_k;ab}=\sum_{c,d=1}^2(g_{\Lambda_k}^{-1/2})_{ab,cd}
\Omega_{\Lambda_k;cd}$$ and next define
$$\Omega_{\Lambda_k\cup\Lambda_{k+1};ab}''=
\Omega_{\Lambda_k;a1}'\otimes\Omega_{\Lambda_{k+1};2b}' -
\Omega_{\Lambda_k;a2}'\otimes\Omega_{\Lambda_{k+1};1b}'.$$ Denote
by $G_{k,k+1}''$ the projector onto the four-dimensional subspace
spanned by $\Omega_{\Lambda_k\cup\Lambda_{k+1};ab}''$ in
$\mh_{\Lambda_k\cup\Lambda_{k+1}}$. A straightforward calculation
shows then that $G_{k-1,k}''$ commutes with $G_{k,k+1}''$. At the
same time, by the property 2) above
\begin{equation}\label{gg}\|G_{k,k+1}''-
G_{\Lambda_k\cup\Lambda_{k+1}}\|=O(\epsilon^l)\end{equation} with
some $\epsilon<1$. Now we use the following abstract observation.

\begin{lemma}
Let $\mh_1,\mh_2,\mh_3$ be three finite-dimensional Hilbert
spaces, and $H_1$ and $H_2$ be two commuting self-adjoint
operators acting on $\mh_1\otimes\mh_3$ and $\mh_3\otimes\mh_2$,
respectively. Then there exists a decomposition
$$\mh_3=\oplus_s(\mh_{3,s}^1\otimes\mh_{3,s}^2)$$ such that for
each $s $ $\mh_1\otimes\mh_{3,s}^1\otimes\mh_{3,s}^2$  and
$\mh_{3,s}^1\otimes\mh_{3,s}^2\otimes\mh_2$ are invariant
subspaces of $H_1$ and $H_2$, respectively, and, furthermore, the
restriction $H_1|_{\mh_1\otimes\mh_{3,s}^1\otimes\mh_{3,s}^2}$ is
an operator acting only on $\mh_1\otimes\mh_{3,s}^1$, and the
restriction $H_2|_{\mh_{3,s}^1\otimes\mh_{3,s}^2\otimes\mh_2}$ is
an operator acting only on $\mh_{3,s}^2\otimes\mh_2$.
\end{lemma}

\begin{proof}
Decompose $H_1=\sum_iH_{11,i}\otimes H_{13,i}$, where $H_{11,i}$
are linearly independent operators on $\mh_1$, and $H_{13,i}$ are
operators on $\mh_3$. Consider the algebra $\mathcal
A_1\subset\mathcal B(\mh_3)$ generated by the operators $H_{13,i}$
and the unity. $\mathcal A_1$ is a von Neumann algebra (closed
under taking adjoints). Similarly, we decompose
$H_2=\sum_iH_{23,i}\otimes H_{22,i}$, where $H_{22,i}\in\mathcal
B(\mh_2)$ are linearly independent and $H_{23,i}\in B(\mh_3)$. By
the commutativity assumption the operators $H_{23,i}$ lie in the
commutant $\mathcal A_1'$ of $\mathcal A_1$. But it follows from
the well-known classification of finite-dimensional von Neumann
algebras (see e.g. \cite{BR}) that there exists a decomposition
\begin{equation}\label{mh3}
\mh_3=\oplus_s(\mh_{3,s}^1\otimes\mh_{3,s}^2)\end{equation} such
that $$\mathcal A_1=\oplus_s(\mathcal B(\mh_{3,s}^1)\otimes{\bf
1}_{ \mh_{3,s}^2}),\quad \mathcal A_1'=\oplus_s({\bf
1}_{\mh_{3,s}^1}\otimes \mathcal B(\mh_{3,s}^2)).$$ Therefore
(\ref{mh3}) is the desired decomposition.
\end{proof}

We apply this observation to the Hilbert spaces
$\mh_{\Lambda_{k-1}},\mh_{\Lambda_{k}},\mh_{\Lambda_{k+1}}$ and
operators $G_{k-1,k}'', G_{k,k+1}''$. The relevant decomposition
then is
\begin{equation}\label{dec}\mh_{\Lambda_{k}}=\mathcal
F_k^1\otimes\mathcal F_k^2\oplus (\mh_{\Lambda_{k}}\ominus\mathcal
G_{\Lambda_k}).\end{equation} Here $\mathcal F_k^1,\mathcal F_k^2$
are two-dimensional Hilbert spaces such that $\mathcal
F_k^1\otimes\mathcal F_k^2=\mathcal G_{\Lambda_k}$. One can choose
orthonormal bases $\{v^1_{k,a}\}_{a=1,2}$ and
$\{v^2_{k,b}\}_{b=1,2}$ in $\mathcal F_k^1$ and $\mathcal F_k^2$
so that $\Omega'_{\Lambda_k;ab}=v^1_{k,a}\otimes v^2_{k,b}.$ The
subspace $\mh_{\Lambda_{k-1}}\otimes\mathcal G_{\Lambda_k}$ is
invariant for $G_{k-1,k}''$, and in this subspace $G_{k-1,k}''$
essentially acts only on $\mh_{\Lambda_{k-1}}\otimes\mathcal
F_k^1$; similarly, $\mathcal G_{\Lambda_k}
\otimes\mh_{\Lambda_{k+1}}$ is invariant for $G_{k,k+1}''$, and it
acts there only on $\mathcal F_k^2\otimes\mh_{\Lambda_{k+1}}$. The
subspaces
$\mh_{\Lambda_{k-1}}\otimes(\mh_{\Lambda_{k}}\ominus\mathcal
G_{\Lambda_k})$ and $(\mh_{\Lambda_{k}}\ominus\mathcal
G_{\Lambda_k})\otimes\mh_{\Lambda_{k+1}}$ lie in the kernels of
$G_{k-1,k}'', G_{k,k+1}''$, respectively (hence we don't factor
$\mh_{\Lambda_{k}}\ominus\mathcal G_{\Lambda_k}$ in (\ref{dec})).

If we use the decomposition (\ref{dec}) for two neighboring blocks
$k$ and $k+1$, we get a decomposition of
$\mh_{\Lambda_{k}}\otimes\mh_{\Lambda_{k+1}}$ as a direct sum of
four subspaces:
\begin{eqnarray*}\mh_{\Lambda_{k}}\otimes\mh_{\Lambda_{k+1}} & = &
(\mh_{\Lambda_{k}}\ominus\mathcal
G_{\Lambda_{k}})\otimes(\mh_{\Lambda_{k+1}}\ominus\mathcal
G_{\Lambda_{k+1}}) \\ & \oplus & (\mh_{\Lambda_{k}}\ominus\mathcal
G_{\Lambda_{k}})\otimes \mathcal F_{k+1}^1\otimes\mathcal
F_{k+1}^2
\\ & \oplus & \mathcal F_k^1\otimes\mathcal
F_k^2\otimes(\mh_{\Lambda_{k+1}}\ominus\mathcal G_{\Lambda_{k+1}})
\\ & \oplus & \mathcal F_k^1\otimes\mathcal F_k^2\otimes \mathcal
F_{k+1}^1\otimes\mathcal F_{k+1}^2.
\end{eqnarray*}
The first three subspaces lie in the kernel of $G_{k,k+1}''$,
whereas in the fourth $G_{k,k+1}''$ acts as the projector onto the
vector
\begin{equation}\label{vk}v_{k,k+1}\equiv v^2_{k,1}\otimes
v^1_{k+1,2}-v^2_{k,2}\otimes v^1_{k+1,1}\end{equation} in the
space $\mathcal F_k^2\otimes \mathcal F_{k+1}^1$.

In order to get a classical model in the sense of Introduction we
introduce additional Hilbert spaces $\mathcal F_k^3,\mathcal
F_k^4$, with $\dim\mathcal F_k^3=\dim\mathcal F_k^4=3^{l/2}-2$
(assuming that $l$ is even), so that $$\mh_{\Lambda_{k}}=(\mathcal
F_k^1\oplus\mathcal F_k^3)\otimes(\mathcal F_k^2\oplus\mathcal
F_k^4).$$ Now define new Hilbert spaces $\ov\mh_{k,k+1}$ by
$$\ov\mh_{k,k+1}=(\mathcal F_k^2\oplus\mathcal
F_k^4)\otimes(\mathcal F_{k+1}^1\oplus\mathcal F_{k+1}^3).$$ The
initial scaled spin chain with sites indexed by $k$ and Hilbert
spaces $\mh_{\Lambda_k}$ assigned to $k$ is then equivalent to the
chain with sites indexed by pairs $(k,k+1)$ and Hilbert spaces
$\ov\mh_{k,k+1}$ assigned to the new sites $(k,k+1)$. Let
$h_{k,k+1}$ be the projector in $\ov\mh_{k,k+1}$ onto the
orthogonal complement to the vector $v_{k,k+1}$ introduced in
(\ref{vk}). Consider the operator
$$H_{\Lambda,0}=3l\sum_{k=1}^nh_{k,k+1}.$$ We claim that if $l$ is
large enough (independently of $n$), then this operator is the
desired classical Hamiltonian, such that the AKLT Hamiltonian
$H_\Lambda^p$ is its relatively bounded perturbation satisfying
assumptions of Theorem 2.

To prove this, we write $H_\Lambda^p$ as
$$H_\Lambda^p=H_{\Lambda,0}+
\sum_{k=1}^n\phi^{(r)}_{k,k+1}+\sum_{k=1}^n\phi^{(b)}_{k,k+1},$$
where $$\phi^{(r)}_{k,k+1}=({\bf 1}-G_{k,k+1}'')\ov H_{k,k+1}
({\bf 1}- G_{k,k+1}'') -l(h_{k-1,k}+h_{k,k+1}+h_{k+1,k+2})$$ will
be the ``purely relatively bounded'' part of the perturbation, and
$$\phi^{(b)}_{k,k+1}=\ov H_{k,k+1}-({\bf 1}-G_{k,k+1}'')\ov
H_{k,k+1} ({\bf 1}-G_{k,k+1}'')$$ the bounded part. First we
estimate $\|\phi^{(b)}_{k,k+1}\|$:
\begin{eqnarray}\|\phi^{(b)}_{k,k+1}\| & = & \|
({\bf 1}-G_{\Lambda_k\cup\Lambda_{k+1}})\ov H_{k,k+1}({\bf 1}-
G_{\Lambda_k\cup\Lambda_{k+1}})\nonumber \\ &  & -({\bf
1}-G_{k,k+1}'')\ov H_{k,k+1} ({\bf 1}-G_{k,k+1}'')\|\nonumber\\ &
\le & \|G_{\Lambda_k\cup\Lambda_{k+1}}-G_{k,k+1}'' \|(\|\ov
H_{k,k+1}({\bf 1}-G_{\Lambda_k\cup\Lambda_{k+1}})\|\nonumber\\ & &
+\|({\bf 1}-G_{k,k+1}'')\ov H_{k,k+1}\|)\nonumber\\ & = &
O(l\epsilon^l),\label{phi2}
\end{eqnarray}
by (\ref{gg}) and because $\|\ov H_{k,k+1}\|\le l$. Now we analyze
the term $\sum_k\phi^{(r)}_{k,k+1}$. We claim that  the condition
(\ref{wcond}) holds with $\alpha=-(1-\gamma/(6l)+O(\epsilon^l))$,
uniformly for all $I\subset \{1,2,\ldots,n\}$. Indeed, note first
that \begin{equation}\label{hg}h_{k,k+1}\le {\bf 1}-
G_{k,k+1}''\le h_{k-1,k}+h_{k,k+1}+h_{k+1,k+2}.\end{equation} It
follows from the right inequality that $\phi^{(r)}_{k,k+1}\le 0.$
Therefore the maximum over $I$ in (\ref{wcond}) is attained when
$I=\Lambda$. By (\ref{sg}),(\ref{gg}) and the left inequality in
(\ref{hg}),
\begin{eqnarray*}\lefteqn{({\bf 1}-G_{k,k+1}'')\ov H_{k,k+1} ({\bf 1}-
G_{k,k+1}'')}& & \\ &  & \ge  \gamma/2({\bf 1}-G_{k,k+1}'')({\bf
1}-G_{\Lambda_k\cup\Lambda_{k+1}})({\bf 1}- G_{k,k+1}'') \\ & &
\ge (\gamma/2+O(\epsilon^l))({\bf 1}- G_{k,k+1}'')\\ & & \ge
(\gamma/2+O(\epsilon^l))h_{k,k+1}\end{eqnarray*} and hence
\begin{eqnarray*}\sum_{k=1}^n\phi^{(r)}_{k,k+1} & \ge &
\sum_{k=1}^n\bigl((\gamma/2+O(\epsilon^l))h_{k,k+1}
-l(h_{k-1,k}+h_{k,k+1}+h_{k+1,k+2})\bigr)\\ & = &
-(1-\gamma/(6l)+O(\epsilon^l))H_{\Lambda,0}.\end{eqnarray*} Since
$\phi^{(r)}_{k,k+1}\le 0,$ this proves our claim about relative
boundedness with $\alpha=1-\gamma/(6l)+O(\epsilon^l)$.

Now we apply Theorem 2. We have
$(1-\alpha)^{\varkappa(1+\nu)}=(\gamma/(6l)+O(\epsilon^l))^{2\varkappa}.$
On the other hand, by (\ref{phi2}), the bounded part of the
perturbation is $O(l\epsilon^l)$, which is asymptotically less
than $(\gamma/(6l)+O(\epsilon^l))^{2\varkappa}.$ Therefore for $l$
large enough $H_\Lambda^p$ is a relatively bounded perturbation of
$H_{\Lambda,0}$ so that Theorem 2 is applicable. The conclusion of
Theorem 3 follows now from Theorem 2, because a sufficiently weak
perturbation of the AKLT model remains within the range of
perturbations of $H_{\Lambda,0}$, where Theorem 2 is applicable.

\section*{Acknowledgements} It is a pleasure to thank Tony Dorlas,
Mark Fannes, Yuri Kondratiev, Robert Minlos, Bruno Nachtergaele,
Sergei Pirogov, Joe Pul\'e and Herbert Spohn for stimulating
discussions and warm hospitality at UCD, DIAS, KUL, TUM and
Universit\"at Bielefeld. The research was supported by the Irish
Research Council for Science, Engineering and Technology.

\end{document}